\documentclass[sigconf]{acmart}

\acmConference[CHI '24]{Proceedings of the 2024
CHI Conference on Human Factors in Computing Systems}{May 11--16, 2024}{Hawai'i, USA}

\acmBooktitle{Proceedings of the 2024 CHI Conference on Human Factors in
Computing Systems (CHI '24), May 11--16, 2024, Hawai'i, USA}

\usepackage{url}
\usepackage{graphicx}
\usepackage{subfigure}
\usepackage{color}
\usepackage{bbding}
\usepackage{soul}
\usepackage{tabularx}
\usepackage{graphicx}
\usepackage{booktabs} 
\usepackage{multirow}
\usepackage{amsmath}
 
\usepackage{amssymb}
\usepackage{geometry}
\usepackage{blindtext}
\usepackage{graphicx}
\usepackage{indentfirst} 
\usepackage{float}
\usepackage{setspace}
\usepackage{makecell}

\copyrightyear{2024}
\acmYear{2024}
\setcopyright{acmlicensed}\acmConference[CHI '24]{Proceedings of the CHI Conference on Human Factors in Computing Systems}{May 11--16, 2024}{Honolulu, HI, USA}
\acmBooktitle{Proceedings of the CHI Conference on Human Factors in Computing Systems (CHI '24), May 11--16, 2024, Honolulu, HI, USA}
\acmDOI{10.1145/3613904.3642959}
\acmISBN{979-8-4007-0330-0/24/05}

\begin{document}

\title[Understanding How Short Video and Live-streaming Platforms Empower Ageing Job Seekers in China]{``There is a Job Prepared for Me Here'': Understanding How Short Video and Live-streaming Platforms Empower Ageing Job Seekers in China}

\author{PiaoHong Wang}
\affiliation{%
  \institution{City University of Hong Kong}
  \city{Hong Kong}
   \country{China}}
\email{piaohwang2-c@my.cityu.edu.hk}
\orcid{0009-0008-8393-7101}

\author{Siying Hu}
\affiliation{%
  \institution{City University of Hong Kong}
   \city{Hong Kong}
  \country{China}}
\email{siyinghu-c@my.cityu.edu.hk}
\orcid{0000-0002-3824-2801}

\author{Bo Wen}
\affiliation{%
 \institution{University of Macau}
  \city{Macau}
 \country{China}}
\email{bowen@um.edu.mo}
\orcid{0000-0003-2287-473X}

\author{Zhicong Lu}
\affiliation{%
  \institution{City University of Hong Kong}
    \city{Hong Kong}
  \country{China}}
\email{zhiconlu@cityu.edu.hk}
\orcid{0000-0002-7761-6351}

\begin{abstract}

\noindent In recent years, the global unemployment rate has remained persistently high. Compounding this issue, the ageing population in China often encounters additional challenges in finding employment due to prevalent age discrimination in daily life. However, with the advent of social media, there has been a rise in the popularity of short videos and live-streams for recruiting ageing workers. To better understand the motivations of ageing job seekers to engage with these video-based recruitment methods and to explore the extent to which such platforms can empower them, we conducted an interview-based study with ageing job seekers who have had exposure to these short recruitment videos and live-streaming channels. Our findings reveal that these platforms can provide a job-seeking choice that is particularly friendly to ageing job seekers, effectively improving their disadvantaged situation.
\end{abstract}



 \maketitle

\section{Introduction}


\noindent The number of unemployed people across the globe has exceeded 235 million in 2020 \cite{Neill}, and job-seeking has been increasingly challenging globally. Additionally, individuals aged forty-five and above, commonly known as the ageing workforce as defined by the World Health Organization (WHO) \cite{chan1997ageing}, often encounter additional obstacles in developing countries such as less interest from potential employers \cite{axelrad2016employers}. In China, these hurdles primarily manifest as a lower rate of job application success and limited opportunities \cite{Discirmination,hu2021age,yeung2021perceived,luyi2021study,Discirmination} for ageing job seekers, placing them in a disadvantaged position within the job market.

In response to the challenges faced by job seekers, the growth of the Internet industry has led to the emergence of various specialized digital platforms for job-seeking, such as LinkedIn \cite{hosain2020linkedin}. These platforms have become valuable sources of employment information \cite{dillahunt2021examining, russomanno2019social} and assisted users in discovering and pursuing prospective job opportunities \cite{rosoiu2016recruiting}. However, ageing individuals, particularly in China, may encounter difficulties in utilizing these specialized job-seeking digital platforms due to limited levels of digital literacy \cite{zhang2022measuring, Xinhua}. Public statistics indicate that only nearly 0.8\% of users on popular specialized job-seeking platforms are over 45 years old \cite{pp}.

In 2022, short videos and live-streams focusing on recruiting ageing workers started to appear on popular short video and live-streaming platforms such as \emph{TikTok} and \emph{KUAISHOU} \cite{kuaishou, pp}. These two platforms have been the most popular short video and live-streaming platforms for Chinese ageing people in recent years \cite{Penta_2020, 199IT_2021,lu2018you,lu2019feel}.
According to a report published by \emph{KUAISHOU}, nearly 20\% of individuals seeking jobs through \emph{KUAISHOU} are over 50 years old \cite{kuaishou}. This figure is several times higher than that of specialized online job-seeking apps.

HCI scholars have long been exploring how information and communication technologies (ICTs) can better enhance the quality of life for the ageing population \cite{colombo2015access,vines2015age,brandt2010communities,righi2017we,ambe2019adventures,yu2023history,knowles2019hci,foverskov2011super}. Their focus primarily centers on addressing age-related limitations, such as reduced social connectedness and physical decline \cite{sun2014being,wilson2022co,piper2013audio,pandya2018taptag}. However, the difficulty that ageing people face in seeking jobs, which naturally increases with age \cite{Discirmination,hu2021age,yeung2021perceived,luyi2021study,Discirmination}, is often overlooked. Most of the prior work concerning digital job-seeking platforms focused on younger users who were proficient in using digital platforms \cite{garg2018or,smith2015searching,dillahunt2016massive} instead of disadvantaged ageing individuals with limited digital literacy \cite{zhang2022measuring,Xinhua}. In contrast to specialized job-seeking platforms, short video and live-streaming platforms are not initially designed for job-seeking but have gained great popularity among ageing job seekers \cite{kuaishou,pp,lu2019vicariously}. However, there is a notable lack of exploration into the lived experiences of ageing job seekers on such platforms.

\begin{figure*}
    \centering
    \includegraphics[scale=0.3]{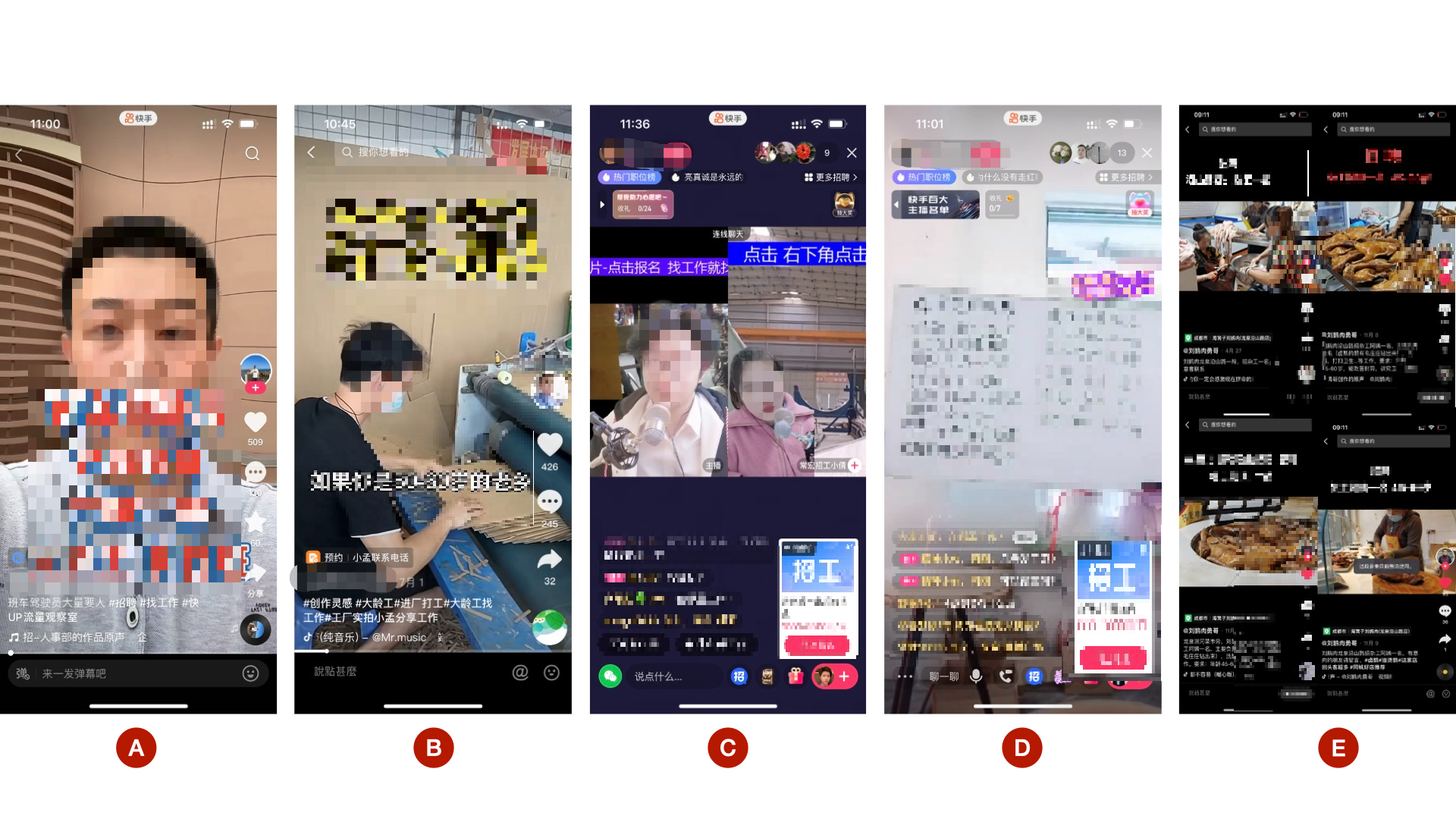}
    \caption{Different job position presentation forms: A) Narrating in front of the camera; B) On-the-Spot position presentation; C) Job introduction through video chat; D) Multiple jobs presented simultaneously; and E) Continuously Updated Videos.}
\label{strategy} 
\Description[A,B,C,D are four figures describing different job presentation forms]{A,B,C,D are four figures describing different job presentation forms. Figure A shows a man is speaking in front of the camera, figure B shows a man is working in the factory, figure C shows two people are conducting video chat in the live-streaming channels, and figure D shows a series short videos concerning recruitment for one position.}
\end{figure*}

With the global ageing population rapidly expanding from 1.4 billion in 2020 to an estimated 2.1 billion by 2050 \cite{World_1}, ageing job seekers deserve the HCI community's attention as they represent a substantially disadvantaged group. Understanding the motivations and practices of ageing job seekers on short video and live-streaming platforms is crucial for the HCI community to reevaluate how to design digital platforms that are inclusive and friendly to ageing individuals, truly benefiting the growing population rather than leaving them behind. Thus, the current research aims to address the following research questions:

\begin{itemize}

    \item \textbf{RQ1:} What are the content and presentation forms of recruitment short videos and live-streaming channels targeting ageing job seekers?

    \item \textbf{RQ2:} How do ageing job seekers utilize short video and live-streaming platforms for job-seeking? 
    
    \item \textbf{RQ3:} How can short video and live-streaming platforms empower ageing job seekers?

\end{itemize}

We conducted an interview-based study with 14 ageing job seekers in China to answer these questions. Several novel and thought-provoking findings are worth enumerating here: i) For ageing job seekers, short video and live-streaming platforms serve as job-seeking avenues that cater to their needs. These platforms mitigate the challenge of limited digital literacy and provide vivid presentations of job opportunities, ii) These platforms can also safeguard the fundamental welfare and rights of ageing job seekers by monitoring the qualifications and public commitments (e.g., monthly wages) made by recruiters targeting ageing workers in their videos, and iii) The content focused on recruiting ageing workers on these platforms mitigates the disadvantages faced by ageing job seekers, enabling them to discover job opportunities that are more accommodating to their physical conditions and daily schedules. 

Our work fills an important gap in HCI by providing novel insights into how ageing job seekers, who are treated unfairly in the labour market and have limited access to digital resources or training, perceive and utilize new platforms for job-seeking. Moreover, our work is situated within the unique social and technological context in China, where the use of technologies among ageing job seekers is understudied in HCI. The findings about the variety of experiences encountered by ageing job seekers can enable platform designers to address potential issues and improve the job-seeking experiences of ageing job seekers. In a nutshell, this research unpacks the potential impact of live-streams and short videos on disadvantaged groups in a non-Western context. 
\section{Related work} 

\noindent In this section, we first review the work on ICT use and acceptance for ageing people and then review the prior work on technology support for recruitment and job seeking in HCI.

\subsection{Digital Platforms Use and Acceptance for Ageing People}

\noindent Researchers in HCI have been exploring 
how information and communication technologies (ICT) could better help the ageing people improve their life quality \cite{colombo2015access,vines2015age,brandt2010communities,righi2017we,ambe2019adventures,yu2023history,knowles2019hci,foverskov2011super}. In 2019, $73\%$ of Americans who aged 65 and older were internet users \cite{livingston2019americans}. In China, 26.3\% of individuals over 50 were Internet users \footnote{http://www.cac.gov.cn/2021-02/03/c\_1613923423079314.htm}. In 2016, Rosales et al. conducted a study concerning the interests of old people using mobile phones and found that older people could be 
active smartphone users for diverse personal interests \cite{rosales2016smartphones}. In HCI fields, technologies designed for the ageing always focuses on addressing limitations due to their age, such as reduction in social connectedness and decline in physical condition \cite{sun2014being,wilson2022co}. For example, Marie et al. designed photographs with audio narrations to facilitate communication between an older adult and others \cite{piper2013audio}, Waycott et al. studied how shared photographs could better help build relationships among older adults \cite{waycott2013older} and Pandya et al. studied how to enhance the learnability for older adults on touch screens  \cite{pandya2018taptag}. However, Lindley et al. claimed the simple assumptions mentioned above may uncover the complex needs of ageing and stresses the importance of independence for ageing people \cite{lindley2008designing}. Similarly, Hornung et al. stressed the extra security and privacy needs of the ageing ICT users \cite{hornung2017navigating}. Along with the development of online services, ageing people also start to spontaneously access online or remote resources out of diverse ICT-related needs \cite{hernandez2009ict,sun2014being}. For example, Chinese ageing match-seekers are trying to match their potential partners through live-streaming channels \cite{he2023seeking} and Pang et al. found ageing people appreciated remote support for their health management \cite{pang2021technology}. However, previous research also found that distrust may be one significant reason for some ageing people to avoid accessing online services \cite{knowles2018older,vines2011eighty,zulman2011trust}.
For example, Knowles et al. found that ageing people may be worried about how their data was used \cite{knowles2018older}, and Zulman et al. found that ageing people may suspect online health information \cite{zulman2011trust}. Therefore, building trust between the ageing and digital platforms is critical for successful ageing ICT technologies \cite{schomakers2018attitudes,sun2014being}. 

Prior studies have found that ageing people are less experienced and cautious in adopting new technologies \cite{schomakers2018attitudes} and the ways in which seniors attend to ICT are shaped by their life attitudes, value systems, and historical specifics \cite{sun2014being}. Prior researchers also identify factors, such as usability, affordability, independence, experience and confidence, as determinants of ageing people's adaptation to specific technology \cite{lee2015perspective}. However, most prior work ignores ageing people's significant needs of job opportunity seeking through digital platforms. It is still unknown what difficulties they may meet during the online job seeking process. Our work revealed that ageing job seekers may have difficulty learning specialized job seeking applications. In contrast, the short video and live-streaming platforms provide reliable and friendly access for them to conduct job-seeking. Our work also provides a nuanced understanding of how ageing people try to identify and interact with potential employers through digital platforms, which provides implications for designing more ageing friendly job-seeking digital platforms.


\subsection{Technology Support For Recruitment and Job-Seeking}

\noindent Recruitment and employment have always been important topics in HCI \cite{chow2013gamifying,delecraz2022making,koivunen2019understanding}. According to previous studies \cite{garg2018or,smith2015searching,dillahunt2016massive}, technology (e.g., social media and automatic systems) has played a significant role in the recruitment and employment process. There has been a lot of studies about recruitment on social media, such as the evaluation indicator of recruitment effectiveness \cite{amadoru2016evaluating}, the strategies and best practices for integrating social media and recruitment activities  \cite{headworth2015social,madia2011best,lu2021uncovering} and the influence brought by social media recruitment \cite{aggerholm2018social,benedict2019recruitment}. More specifically, Ari et al. studied how automatic systems can be applied to the faculty recruitment process and help the college save time and cost \cite{visa2015new}, and Leung et al. studied the potential bias brought by these automatic recruitment systems \cite{leung2020race}. 

Not only for recruiters, but technology also benefits job seekers, especially marginalized groups. For example, Hendry et al. studied how peer-to-peer sharing system match the young homeless to potential job positions \cite{hendry2017homeless}. Previous researchers also indicate that it is hard for disadvantaged groups to make full use of the technology to support their job seeking process \cite{wheeler2018navigating}. Similarly, Ihudiya et al. found that returning citizens can hardly extend their knowledge of digital technology to support job-searching tasks \cite{ogbonnaya2018returning}. Therefore, researchers are dedicated to designing more minority-friendly systems to support disadvantaged job seekers \cite{dillahunt2021implications,dillahunt2016designing}. For example, Suzuki et al. designed a micro-internship platform that can help workers who cannot afford to invest their earnings in learning a skill \cite{suzuki2016atelier}, and Tawanna et al. designed a system to help low-resource job seekers identify the essential skills needed for their dream jobs \cite{dillahunt2019dreamgigs}. Tawanna et al. also proposed ten tangible system design concepts to address the needs of underserved job seekers, which mainly include special designs helping job seekers receive timely feedback as well as supporting their job-seeking process \cite{dillahunt2018designing}. 

Although HCI scholars have made significant efforts to enhance support for marginalized job seekers, there is little relevant research in HCI to consider the use of social media for disadvantaged job seekers \cite{ogbonnaya2018returning,lu2021uncovering}. Ihudiya et al. studied how returning citizens interact with digital technologies for job searching and tried to observe their practices and attitude towards social media (e.g., Facebook). Unfortunately, the interviewees they recruited expressed a strong avoidance of social media, which led to a lack of observation of social media use by these returning citizens. Instead, Ihudiya et al. reported the daily technology (e.g., mobile phones) they used. Conversely, Lu et al. studied the job-seeking behaviour of marginalized groups in social media \cite{lu2021uncovering}. Lu et al. described their practices from the perspective of recruiters instead of job seekers. They reported how employers could recruit low-wage job seekers from Facebook and the behaviours exhibited by these low-wage job seekers. While there have been many efforts in HCI that addressed the job searching needs of diverse disadvantaged groups including returning citizens \cite{ogbonnaya2018returning} and marginal social media platform users \cite{lu2021uncovering}, no studies so far offer an empirical investigation of how Chinese ageing job seekers utilize short video and live-streaming platforms as well as how these platforms may empower them during their job-seeking process. These platforms could provide valuable opportunities for HCI scholars to understand how to design ageing-friendly job-seeking systems. Our work fills this important gap by providing a nuanced understanding of experiences and practices of ageing job seekers on such platforms.


\section{Background: Chinese Ageing Job Seekers}

\noindent According to the World Health Organization (WHO), the proportion of ageing workers who are over 45 years old in the labour market is expected to rise to 41.3\% in 2055 \cite{chan1997ageing}. These ageing workers have to face daily deterioration in physical condition and adaptability to the job task, which may cause a decline in their job performance and make it hard for them to be steadily employed \cite{chan1997ageing, mcdaniel2012job}. Long-term unemployment brings severe consequences to individuals, such as health and financial problems \cite{zacher2013older,frese1987prolonged,klehe2012ending,ranzijn2006scrap}. For example, researchers in Chile have found that the health of unemployed ageing workers is significantly worse than that of employed ones \cite{vives2018gender}. The Chinese labour participation rate was 56.1\% for women and 68.2\% for males between the ages of sixty and sixty-nine in 2015 \cite{henry2018drives}, which means more than half of the elderly people, even those over the retirement age still need a job to support themselves. However, researchers indicate that these ageing unemployed people face the greatest difficulty in being reemployed \cite{difficult}. 

Age discrimination is already a common and serious social phenomenon in China's job market. Zhou et al. confirmed the widespread age discrimination in the Chinese labour market through an empirical study of about 300,000 job advertisements in the cities of Shanghai and Chengdu, which mainly includes unreasonable age restrictions imposed by recruiters on the job positions \cite{Discirmination}. Liu et al. proposed that the ageing job seekers are also prone to being treated unfairly by recruiters even after being successfully employed, which can be unequal treatment in promotion, performance appraisal, and wages \cite{Unfair}. Our work provides an HCI lens to understand how ageing job seekers utilize short video and live-streaming platforms to conduct job seeking as well as how these platforms could empower them during the job seeking process. Surprisingly, we found that video entertainment platforms can significantly assist ageing Chinese job seekers in their job-seeking efforts and in enhancing their overall circumstances. This enhancement encompasses better safeguarding of their fundamental rights and welfare. 

\section{Method}

\noindent Inspired by Tang's research, which utilized live-streaming channel observation and streamer interviews to analyze how live streaming empowers Chinese rural women \cite{tang2022dare}, we employed the same framework for data collection and analysis in our study. Initially, we observed 100 live-streaming channels (comprising 50 channels from \emph{TikTok} and 50 from \emph{KUAISHOU}), as well as 100 short videos (50 from \emph{TikTok} and 50 from \emph{KUAISHOU}) to address RQ1. Subsequently, we conducted interviews with 14 ageing job seekers to delve into their motivations, practices, and experiences related to job-seeking on these video entertainment platforms. The study protocol was approved by our institutional review board (IRB). We started our interview and recording only after we have the interviewees' consent. During the interviews, they had the right to refuse to answer any questions or to terminate the interviews. Our data collection process specifically encompassed two stages:

\begin{itemize}

\item[-] \textbf{Step 1. Observing Live-streaming Channels and Short Videos}: We observed 100 ageing worker recruitment short videos and 100 ageing worker recruitment live-streaming channels on \emph{TikTok} and \emph{KUAISHOU}. We identified these videos and channels based on their recruitment-related content and their age requirement.

\item[-] \textbf{Step 2. Interview with Ageing Job Seekers}: In this phase, we conducted semi-structured interviews with 14 ageing job seekers who were actively seeking employment opportunities through \emph{TikTok} and \emph{KUAISHOU}. It is worth noting that, in alignment with the definitions of the ageing workforce established in prior literature \cite{chan1997ageing}, all our interviewees were aged 45 or older.

\end{itemize}

\begin{table}[]
\caption{Summary of ageing job seekers interviewed.}
\small
\begin{tabular}{ccccc}
\toprule
ID & Age & Sex & Position Applied           & Educational Level  \\ \midrule
1  & 46  & F   & Factory Cleaner            & High School        \\
2  & 50  & M   & Car Factory Worker         & Unrevealed         \\
3  & 50  & F   & Cleaner                    & Associate Degree   \\
4  & 50  & F   & Home Craftswomen           & Primary school     \\
5  & 54  & M   & Machine Factory Worker     & Primary school     \\
6  & 55  & F   & Textile Factory Worker     & Primary school     \\
7  & 56  & M   & Electronics Factory Worker & Primary school     \\
8  & 57  & F   & Restaurant Waitress        & Primary school     \\
9  & 58  & M   & Rubber Factory Worker      & Primary school     \\
10 & 58  & F   & Hotel Cleaner              & Junior high school \\
11 & 45  & F   & Home Craftswomen           & Primary school     \\
12 & 45  & M   & Restaurant Chef            & High school        \\
13 & 52  & F   & After-school teacher       & High school        \\
14 & 46  & F   & Nurse                      & Primary school     \\ \bottomrule
\end{tabular}
\Description[The demographics of our interviewees]{This table describes the age, sex, position Applied, and educational levels of our interviewees as well as the corrsponding positions they applied for. There are five male inteviewees and nine female interviewees. Most of them did not enter universities. Most of the jobs they applied for were workers. Their ages range from 45 to 58.}
\end{table}

\subsection{Observing ``Ageing Worker Recruitment'' Short Videos and Live-streaming Channels}

\noindent To gain a comprehensive understanding of the short videos and live-streaming channels targeting ageing job seekers, our primary focus was on content that contained the keyword "ageing worker recruitment" in the titles or labels. We selected \emph{TikTok} and \emph{KUAISHOU} as our observation platforms since these two platforms have been the most popular short video and live-streaming platforms among Chinese ageing users \cite{Penta_2020, 199IT_2021}. In total, we reviewed 100 short videos and 100 live-streaming channels related to "ageing worker recruitment." During the observation phase, we captured screenshots of key moments, such as video covers (e.g., the cover of the videos in \autoref{strategy}.a). Additionally, we closely monitored interaction activities between ageing job seekers and recruiters, including comments from job seekers and responses from recruiters, for the observed videos and live-streaming channels. We also examined any supporting plugins provided by the platforms, such as resume generators on \emph{KUAISHOU}, and made detailed notes regarding their functionality. Subsequently, we merged the material obtained from our observations with the transcripts from the subsequent interviews. This combined dataset was then subjected to open coding analysis \cite{corbin2008techniques}. Throughout the entire observation process, our primary focus was on the following aspects:


\begin{itemize}

    \item {I)} How do the recruiters present the job positions in the live-streaming channels and short videos? This can help us have a basic understanding of the content of these videos and channels.              
    \item {II)} How do the ageing job seekers interact with recruiters through short videos and live-streaming channels? Common comments posted by job seekers include questions about the job salary and welfare, such as \emph{`Can the factory provide insurance to us?'}, which help us understand the practice of ageing job seekers.
    
    \item {III)} What technological assistance can short videos and live-streaming platforms provide to these ageing job seekers (e.g., the plugin for application submission on the live-streaming channel interface)? This can help understand how short videos and live-streaming may empower ageing job seekers.
            
\end{itemize}

\subsection{Interviews with Ageing Job Seekers on Short-video and Live-streaming Platforms}

\noindent To further understand the practical experience of ageing job seekers on \emph{TikTok} and \emph{KUAISHOU}, we conducted a qualitative semi-structured interview with 14 job seekers who have watched these videos and channels. Before starting the formal interview, we confirmed with our interviewees that they had successfully found jobs or were trying to find jobs through short videos and live streaming channels, as they were all over 45 years old. We recruited six interviewees by sending direct private messages to job seekers who commented under the related videos or on live-streaming channels. Besides, we joined the chat group created by recruiters and recruited two additional interviewers. We seek four additional interviews using the personal connections of these interviewees (e.g., their friends) and the interviewee advertising poster posted on social media (e.g., \emph{TikTok}) by the first author. We conducted the semi-structured interview between September 2022 and September 2023. All interviewees are paid between 100 and 200 Chinese Yuan. Due to equipment and network issues, some interviewees answered our questions via text with the help of their young family members. The interviews were conducted in Chinese and recorded by the first author.

\subsection{Data Analysis}

\noindent As mentioned in section 4.1 and 4.2, the data we analyzed was made up of three parts: i) transcripts of the interviews, ii) notes of observed short videos and live-streaming channels, and iii) screenshots of observed short videos and live-streaming channels. We
conducted interpretive qualitative analysis for our text data ( i and ii ) \cite{merriam2019qualitative}.
Inspired by Xiao \cite{xiao2020random},we also started with open coding in two phases \cite{charmaz2006constructing}.
In the first phase, we coded our transcripts line-by-line. Examples of such codes include ``low difficulty'' and ``automatic recommended videos''. In the second phase, we synthesized the codes from the previous phase to extract higher level themes. Examples of these higher themes include ``job-seeking process'', ``content of recruitment videos and live-streaming channels'' and ``empowerment for ageing job seekers.'' Inspired by Xiao \cite{xiao2020random}, we also used the screenshots collected during the observations as supplementary data to our transcripts and developed codes for them. The codes for screenshots described the content of videos and live-streams for instance: ``narrating in front of the camera'' and ``on-the-spot presentation.'' During the interview, we found that interviewees frequently mentioned their negative experience with specialized job-seeking apps. Although this information is not directly related to short video and live-streaming platforms, we still consider it as part of the motivations for respondents’ job-seeking on short video and live-streaming platforms.

\subsection{Ethical Considerations}

\noindent First, all our interviewees have used \emph{TikTok} and \emph{KUAISHOU} just for fun before conducting job-seeking on them. Therefore, readers should not be misled into thinking our interviewees were forced to access these platforms in order to conduct job-seeking. To protect the privacy of our interviewees, we also de-identified some of their personal privacy information (e.g., birthplace) before conducting our data analysis.

\section{FINDINGS}

\noindent In this section, we will begin by introducing the diverse presentation forms of observed short videos and live-streaming channels. Then we will introduce the live experiences of ageing job seekers on these platforms as well as the empowerment brought by these platforms.

\subsection{Content of Recruitment Videos and Live-streaming Channels}

\noindent Throughout our observation process, we identified six primary content elements commonly featured in recruitment videos and live-streaming channels: I) Salary: Recruiters often begin by presenting the salary and benefits associated with the positions being advertised. For instance, they might say, ``Toy Factory Worker in Hei LongJiang! Five thousand yuan monthly, one day off each week, breakfast and lunch provided.'' Based on our observations, most positions appearing on \emph{TikTok} and \emph{KUAISHOU} offer a monthly salary of under 10,000 yuan (approximately 1,380 US dollars), II) Welfare: Alongside salary information, recruiters also described the welfare packages associated with the positions being advertised, such as the provision of daily meals or accommodation, III) Environment: Recruiters provide insights into the work environment. For example, they might showcase the factory where job seekers would be working in, IV) Requirements: Detailed recruitment requirements concerning the positions such as ``You only need to know how to write 26 letters. Anyone under seventy, male or female, can come!'', V) Workplace Location: Information about the location of the workplace, including the province or region where the job is situated, VI) Daily Tasks: The specific tasks that workers would be expected to perform. After conveying this essential information, recruiters may provide contact information as video comments or answer questions posed by job seekers in the live-streaming channels. These questions can range from inquiries about factory management styles to inquiries about regional cultural differences and can sometimes be quite direct or even impolite, such as, ``I have heard that people in your province are rude, is this true? Is the food eaten there in a southern or northern style? '' Recruiters' responses to these questions offer further insights into the positions and contribute to the dynamic and interactive nature of job presentations through videos and live-streaming channels, fostering engagement between recruiters and job seekers.

\begin{figure*}
    \centering
    \includegraphics[width=0.8\linewidth]{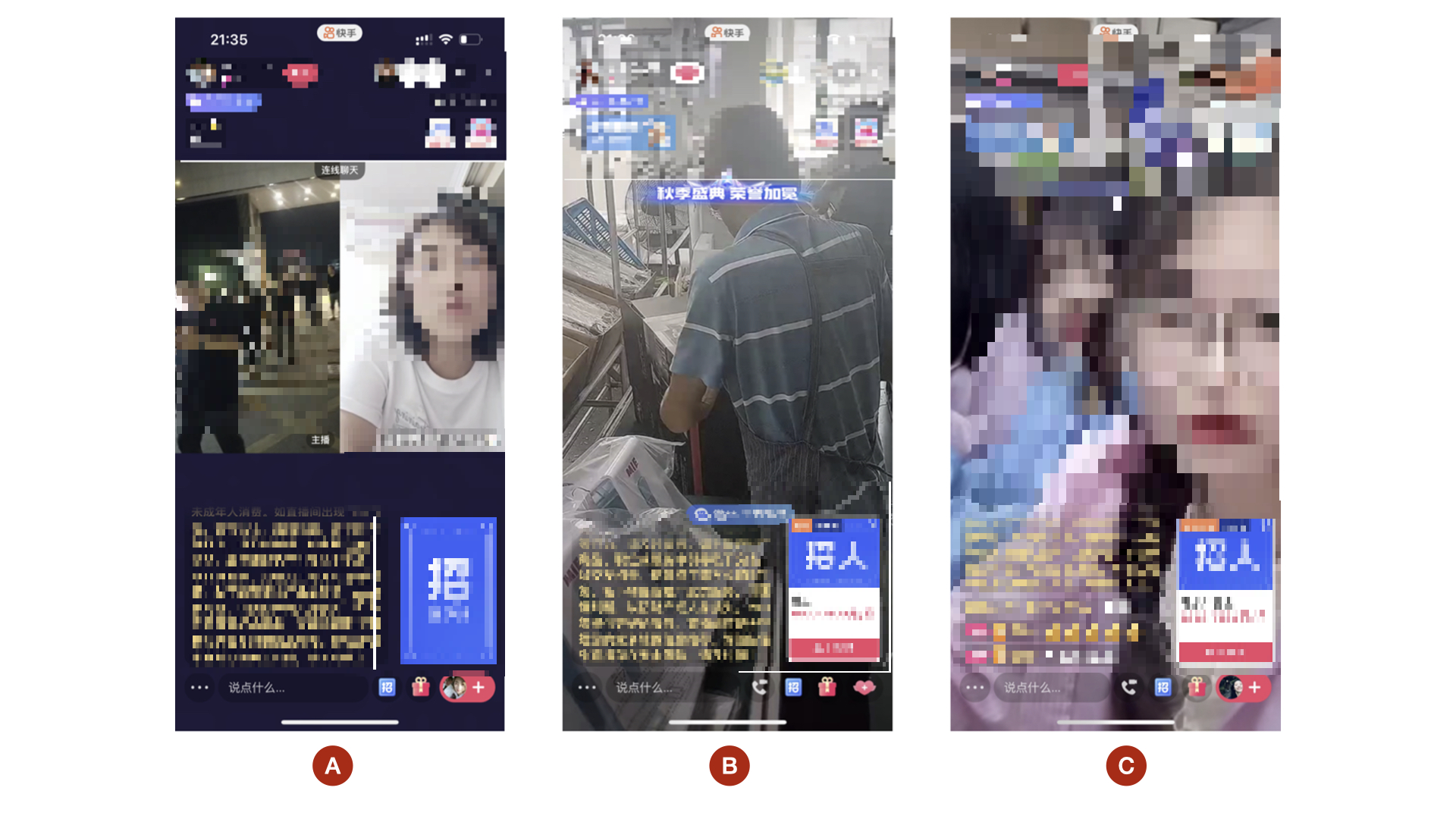}
    \caption{Comparison between there forms of job presentation: A) Video chat, the recruiter was chatting with other people in the factory to introduce the position; B) On-the-spot presentation, the daily tasks of workers and the workspace environment in the factory are shown through the lens; C) Narrating in front of the camera, the recruiters were introducing the position directly in front of the camera.}
\label{detail} 
\Description[A,B,C are three figures describing different job presentation forms]{A,B,C are three figures describing different job presentation forms. Figure A shows a female recruiter is conducting video chat with others in the factory, figure B shows a worker is working in the factory, and figure C shows two female recruiters in front of the camera.}
\end{figure*}

\subsection{\textbf{Job Presentation Forms on Short Video and Live-streaming Platforms}}

\noindent In this section, we will introduce various job presentation forms in these videos and live-streams. These forms range from simple narrations in front of cameras to more dynamic job presentations like on-the-spot job presentation. These diverse presentation forms provide recruiters with a wider range of options to attract potential job applicants and create more engaging job presentations. Statistically, narrating in front of the cameras and on-the-spot position presentation are the two most common job presentation forms, the sum of which accounts for more than 60\% of the observed short videos and live streams. Our interviewees also prefer to learn more about the real working environment from on-the-spot videos and live-streaming channels, which is the most popular job presentation form among our interviewees.

\subsubsection{Narrating in Front of Cameras}

Some recruiters choose newscaster style presentation, where they face the camera and systematically introduce job information step by step (e.g. \autoref{strategy}.a). They may also use factory images as their background or show photos and videos taken from the workplaces. Typically, this type of presentation begins with details about the job positions’ salary and the welfare offered by the companies, followed by a description of the specific daily tasks. To emphasize the attractiveness of the job positions, recruiters may compare them to similar positions available in the labor market in terms of salary and welfare. This presentation form is the most common one of job presentations in both recruitment videos and live-streaming channels. 
\subsubsection{On-the-Spot Position Presentation}

Some recruiters choose to conduct on-the-spot job position presentations. They may present the job presentation in the recruiting factories and walk around to show the workspace environment. Some of them may interview one working employee and ask for details of the job position (e.g., work pressure) from the employees. Compared to regular job position descriptions, this strategy is a more realistic way to show the work environment as well as the workers' daily work. 

\subsubsection{Job Introduction through Video Chat}

Some recruiters engaged in video chats with other management-level staffs, such as the factories' human resource managers, during the live-streaming process. Unlike the two previous job presentation forms, video chats involve multiple individuals in the job introduction process. The recruiters may provide overviews of the jobs, while another key individual provides additional details, offering job seekers a more comprehensive understanding of the positions. This presentation form is more commonly used in live-streaming channels.

\subsubsection{Multiple Jobs Presented Simultaneously}

Some recruiters introduce multiple job positions simultaneously. For instance, they might recruit restaurant chefs, managers, and waiters at the same time, making the recruitment process more efficient and appealing to a broader range of job seekers. This presentation from is common in live-streaming channels, where recruiters can provide personal recommendations to each job seeker who sends real-time comments. Recruiters often reiterate the salary and welfare of advertised positions. 


\subsubsection{Continually Updated Videos for One Position}

Due to the length limitation of short videos on \emph{TikTok} and \emph{KUAISHOU}, some recruiters may take multiple short videos to present a more comprehensive introduction to the positions or answer questions from viewers. It is common for them to introduce the working environment, staff welfare and daily life one by one in a series of short videos. Sometimes, in the case of unexpected events (e.g., city lockdown), recruiters may shoot new short videos to notify job seekers that the recruitment process is on hold, which also improves the efficiency of information delivery.

\subsection{Job-Seeking Process on Short Video and Live-streaming Platforms}

\noindent In this section, we will explore the job-seeking process of ageing job seekers on short video and live-streaming platforms. The process encompasses the selection of jobs from system recommendations, proactive interactions with recruiters via platforms, signing up and self-presentation through platforms, and reaching out to recruiters beyond platforms.



\subsubsection{Selecting from Auto-Recommended Videos and Channels}

\noindent Most interviewees reported to us that their first exposure to recruitment content was through system recommendations. The platforms' recommendation algorithms based on their browsing history allows them to receive extremely accurate job position recommendations. For example, interviewee P12 is a restaurant chef and enjoys watching cooking videos on \emph{TikTok}. Until one day, he received a short recruitment video recommendation for a restaurant in Shanghai. Unlike regular short videos as well as live-streaming users, these ageing job seekers rarely searched for related content proactively through keywords. Instead, they rely heavily on automated system recommendations because of the lack of related technology skills ( As P7 reported \emph{``I don't usually type, and I don't know how to type, I just know how to write with my hands.''}) and the continuation of previous entertainment habits using these platforms. Some interviewees adopted special strategies to get richer recommendations from systems, which include manually adjusting their location information in the systems (e.g., users can select their location cities under the "same cities" section in \emph{TikTok}). Both \emph{TikTok} and \emph{KUAISHOU} allow users to only receive content published by nearby people, which facilitates ageing job seekers in finding job positions across regions. P9 reported to us that he frequently switched his job positioning to different neighbourhoods then he found a local factory of the city that was hiring ageing workers, which is a half-hour drive from his home.  

\subsubsection{Consulting with Recruiters through Chats and Comments}

In recruitment live-streaming channels, recruiters provide detailed descriptions of job positions, including information about the salary package and the working environment. Ageing job seekers utilize the interactive features provided by short videos and live-streaming platforms to engage with recruiters and obtain more relevant information. Within these live-streaming channels, job seekers can send real-time text messages to the recruiter, inquiring about any specific details they are interested in.

Some recruiters may present multiple job positions simultaneously and make recommendations based on the information provided by job seekers in live-streaming channels. For example, a job seeker might inquire, ``I'm from northern China, I'm almost 60 years old, I'm a farmer, and I only work in the fields, do you have any recommendations?'' The recruiter can respond with a recommendation like, ``Yes, you can come and work in the biscuit factory. The biscuit factory doesn't require much skill or strength, and you can just sit and work.'' To protect the privacy of the questioner, recruiters often conceal the names of users who speak in the channels, ensuring that users don't have to worry about their privacy being compromised. For job seekers with limited literacy, the platforms allow them to communicate with recruiters through voice chat, further encouraging interaction. In contrast to live-streaming, job seekers in short videos primarily interact with recruiters through comments on videos and private messages. Several interviewees mentioned that they value the public questions asked by other job seekers in the comments section of videos and the corresponding replies from recruiters. These interactions help them gain a more comprehensive understanding of the job position.   

\subsubsection{Submitting Applications and Conducting Self-Presentation}

The job seekers may contact the recruiter and sign up by sending a private message. However, the platforms also provide special plugin support for the recruitment short videos and live-streaming channels (e.g., plugin \emph{KUAIZHAOGONG} in \emph{KUAISHOU}), which makes it extremely easy for ageing job seekers to apply for job positions in the channels, as all they have to do is to click on the apply button in the bottom of the screen and select the job positions they prefer. This page also displays the responsibilities, basic earnings and welfare for each job position, as well as the qualifications for recruiters. After the final confirmation of the selected job positions, the job seekers can submit the application. The recruiters could contact the suitable applicants via the phone number they register on the platforms. In order to increase the chances of a response, it's common for ageing job seekers to deliberately portray themselves on the platforms to prove their fitness and experience (e.g., uploading working videos and photos to personal pages). For example, P6 changed her avatar from a cartoon character to a her real image and posted working videos in a textile factory after she started seeking opportunities in \emph{TikTok}. Job seekers were usually contacted by phone after sending a private message or submitting their applications. According to the feedback from our interviewees, the KUAIZHAOGONG plugin greatly improves the efficiency of application submission. More than half of our interviewees had used the KUAIZHAOGONG plugin for job-seeking. All of them agreed that the plugin successfully lowered the difficulty of online job-seeking process.

\subsubsection{On-Site Visiting and Accepting the Appointments}

Although the live-streaming channels and short videos, as well as the private WeChat contact, could offer a preliminary introduction to the job positions and working environment, most of our interviewees report that they still need an onsite visit before making final decisions. Unlike employment agencies, job seekers are not required to pay a referral fee in advance before first contact with factories and recruiters in reality. They could physically inspect the working conditions of workers and think twice about whether they are the right person for the positions. Some recruiters offer free accommodation to them and provide medical examinations to ensure that they are competent for the jobs.

\subsection{\textbf{Empowerment for Ageing Job Seekers}}

\noindent In this section, we will describe how short video and live-streaming platforms empower ageing job seekers compared to other job-seeking access. During the interview process, we found that these platforms could provide age-friendly and reliable online access for ageing individuals seeking jobs. These platforms can also assist them in improving their disadvantaged situation in the workplace, including better protection of their basic welfare and rights, as well as alleviating age-related challenges. Most of our interviewees compared different ways of job-seeking based on their own experience and pointed out that job-seeking via short video and live-streaming platforms makes up for certain significant shortcomings of specialized job-seeking platforms and offline job-seeking access.

\subsubsection{Ageing Friendly Online Access to Seek Jobs}
\par
Firstly, the short video and live-streaming platforms successfully provide a low digital friendly job-seeking access to our interviewees, which is also more vivid and reliable compared to other job-seeking access.

\setlength{\parindent}{0cm} \paragraph*{Low Digital Literacy Friendly Job-Seeking Access}

During the interview process, some of our interviewees shared their experiences concerning job-seeking on specialized job-seeking platforms. Compared with specialized job-seeking platforms, most of our interviewees agreed that these short video and live-streaming platforms provided them with more ageing-friendly job-seeking experiences. Consistent with prior studies on the limited digital literacy problem of ageing people \cite{zhang2022measuring}, some of our interviewees expressed their discomfort with these specialized job-seeking platforms. Unlike easy-to-understand short video and live-streaming platforms, it is often difficult for our interviewees to learn and understand specialized job-seeking platforms. For example, P7 could barely scan relevant job position information on specialized job-seeking platforms with the help of her daughter after one week of instruction and P2 can't understand the business model of these specialized platforms. P2 mistakenly believed all job positions in the platforms were from the same recruiter and questioned the fairness of the platforms: \emph{`` Many of the pages on them are bought out by the same labor agency.''} 

\parindent10pt Although most of our interviewees were resistant to specialized job-seeking platforms, they were very skilled in using easy-to-understand short video and live-streaming platforms such as \emph{TikTok} and \emph{KUAISHOU}. One reason for this is that these entertainment apps do not require many complex operations (e.g., typing and searching). They shared with us that they only need to \textbf{\emph{``swipe the videos''}} on TikTok and KuaiShou in their daily lives, which means they can finish the process of selecting and watching the short videos and live-streaming by simply swiping the screen. This greatly helps them avoid the burden of the learning process. More importantly, \emph{TikTok} and \emph{KUAISHOU} have sophisticated recommendation systems, which can automatically recommend suitable job positions to the ageing job seekers based on their browsing history rather than require their proactive operations. 

\setlength{\parindent}{0cm} \paragraph*{Vivid and Reliable Access to Learn the Jobs}

Unlike traditional text-based recruitment information in online recruitment apps or paper-base job advertisements in the offline labour market, most interviewees reported that the recruitment short videos and live-streaming channels were more vivid and provided more reliable detail about the job position for them to make judgement, which can also greatly help them avoid being scammed. This perceived safety successfully motivated our interviewees to conduct job-seeking on these novel platforms. P12 reported to us that he could have a complete and real understanding about the workplace environment, staff daily life and schedule from the recruiters on \emph{TikTok}. Compared to information in recruitment apps, our interviewees also reported that the \textbf{real} recruiters who appear in short videos and live-streaming channels are more likely to be trusted by them, e.g., \emph{``Many of the recruiters on \emph{TikTok} are real people presenting job positions, and there are many female recruiters. I trust female recruiters more.''} (P8). We also found that the perceived safety towards these platforms tend to be based on the trust between interviewees and recruiters instead of the actual safety mechanisms of these platforms, which is also different from other specialized job-seeking platforms. Some of our interviewees tend to trust the recruiters with larger numbers of followers, more uploaded videos and more active replies to their followers. They stressed the importance of interaction between job seekers and recruiters for them to make application decisions, e.g., \emph{"I can justify whether the recruiter is trustworthy based on people's comments.''} (P3). 

\parindent10pt Because of the scarcity of opportunities in daily lives, several of our interviewees have
also tried to find suitable jobs through specialized online job-seeking platforms. Unlike short video and live-streaming platforms, it is often difficult
for them to identify whether the recruiters on these text-based platforms are fraud or not and hardly would platforms review
the recruiter information for them in advance. Several of them have suffered from the recruiters' deception in specialized job-seeking platforms, which can be unrealistic salaries, fake daily task descriptions and even fake workspace locations. Worse still, some recruiters may ask our interviewees to pay various fees in advance without any prior guarantee. As P12 mentioned: \emph{`` The recruiter said the workplace was in Shenzhen, but when I arrive, they admitted that the workplace was actually abroad ... they wanted me to pay a sum of money in advance for an exit permit. I didn't consider going abroad ... I only had a high school education ... if I did I might not be able to go back to my country.''}


\begin{figure*}
    \centering
    \includegraphics[width=0.8\linewidth]{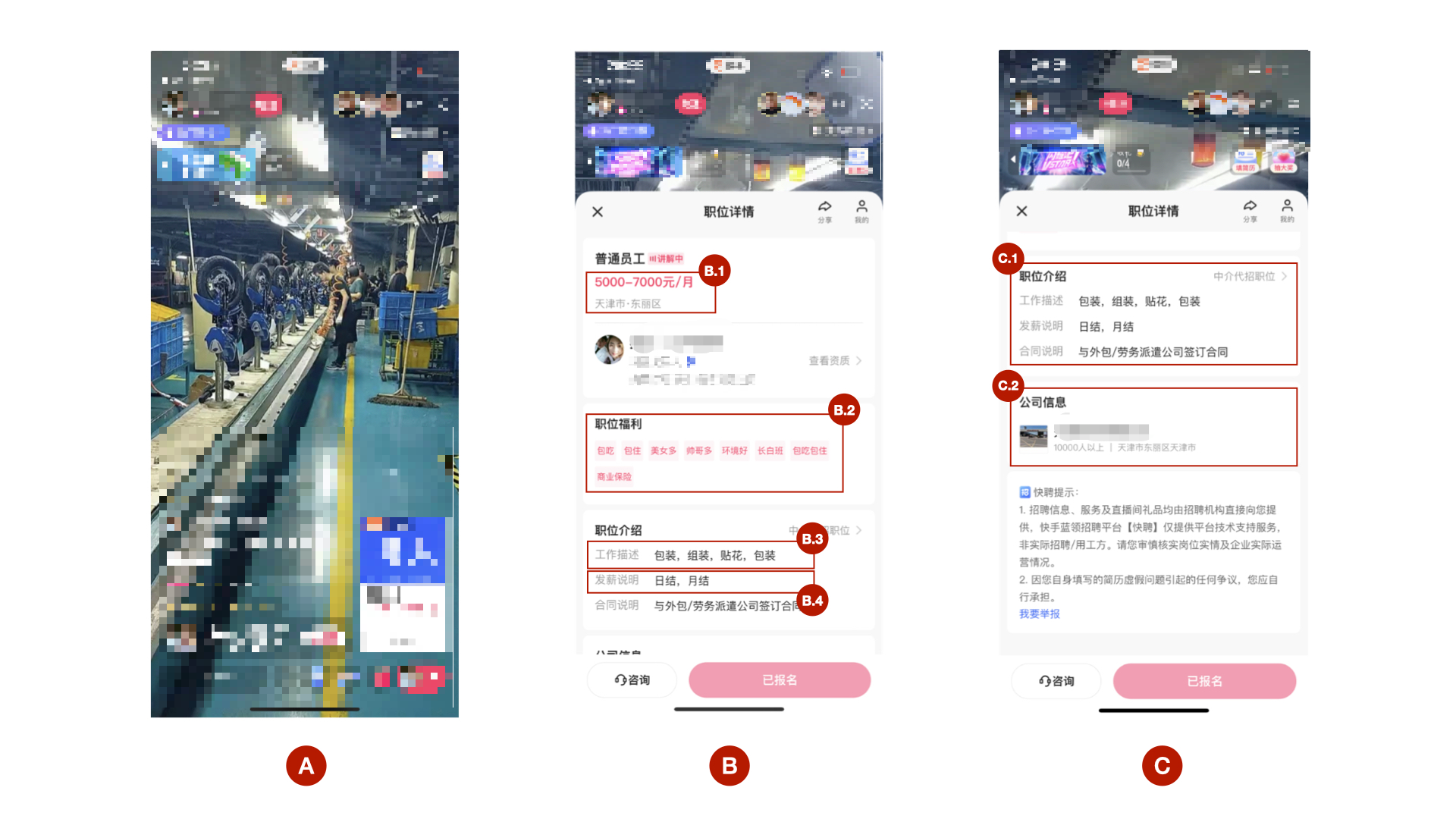}
    \caption{The detailed information concerning the job published in one \emph{KUAISHOU} live-streaming channel through \emph{KUAIZHAOGONG} plug-in, which includes: Monthly Income (B.1), Welfare(B.2), Daily Tasks(B.3), Salary Settlement(B.4), Job Description (C.1) and Company Information of the position (C.2). Job seekers can directly submit applications through this page.}
    \label{detail_2} 
    \Description[A,B,C are three figures describing the job introduction page on KuaiShou Platform]{A,B,C are three figures describing the job introduction page on KuaiShou Platform. Figure A shows the environment of the factory, Figure B shows the monthly salary, welfare and daily task of the position, and Figure C shows the daily task and company information.}
\end{figure*}

\subsubsection{Improving the Situation of Ageing Job Seekers in Workspace} Secondly, the short video and live-streaming platforms also successfully improve the situation of ageing job seekers by better protecting their basic welfare and rights as well as alleviating their disadvantaged situation.

\setlength{\parindent}{0cm} \paragraph*{Better Protecting Basic Welfare and Rights of Ageing People}

Compared with previous job-seeking methods, our interviewees indicate that live-streaming and short videos as new access to conduct job-seeking can better protect their basic welfare and rights. In \emph{KUAISHOU}, the recruiters need to clearly state the employee welfare as well as display their qualification certificates to the ageing job seekers publicly on the platforms (\autoref{detail_2}), which could greatly reduce the ageing workers' risk of unexpected welfare deprivation after entering the companies. Through short recruitment videos and live-streaming channels, some of our interviewees successfully access large companies which are willing to offer them staff bonuses. For example, P2 successfully joined BYD, the world's top 500 car manufacturing company, which was willing to reward him with 10,000 Chinese yuan for each successful applicant he recommended to the company. Some of our interviewees had previously been living in developing regions and had been enduring unreasonable welfare deprivation in small and informal companies for tens of years. For example, P12 reported that he first time enjoyed the basic welfare after entering a restaurant of a catering group in Shanghai following the recommendation from \emph{TikTok} after working for small restaurants in Shanxi province (A province where the main industry is coal mining) for years. Some violations of employees' rights, such as withholding and defaulting on wages of ageing workers, have also been effectively mitigated among job positions recommended by \emph{TikTok} and \emph{KUAISHOU}. The recruiters on these two platforms are required to clearly declare the salary range and the monthly payment date, and most of the interviewees reported that these companies are required to sign formal contracts with them, which greatly protects the rights of ageing job seekers. P7 informed us that he carefully checked details in the contract before entering the electronics factories and focused on anything important affecting his income (e.g., probationary period). As he reported to us \emph{``I would be concerned about the length of the probationary period. I don't want to work for a period of time and then lose my job.''} 

\parindent10pt Different from recruiters who make public commitments on short video and live-streaming platforms, recruiters in the offline labor market often refuse to sign formal contracts with the ageing workers to reduce costs and maliciously deprive ageing workers of their basic welfare. The lack of contracts makes it difficult for ageing workers to protect their rights, and they can only bear the deception from recruiters. As P11 shared her father's experiences to us: 
\begin{quote}
\emph{``My father used to be a construction worker ... His recruiter took him to the construction site ... Every time my father returned home for the Chinese New Year, the recruiter began to default on my father's wages ... Only when the recruiter felt my father was needed again would he be willing to settle the arrears.''} 
\end{quote}
The absence of basic welfare items also results in ageing employees being unable to secure their future, even losing hope for successful retirement. Interviewee P3 reported to us that they need to pay over 100,000 yuan in bills for endowment insurance because none of her former recruiters paid for her. In addition to the basic insurance and funding welfare, some interviewees reported to us that they were also not entitled to welfare such as paid holidays, weekends off and so on, which may make them tired and difficult to work for long periods of time, e.g., 
\begin{quote}

\emph{``I was the only cook in the restaurant. If I left, the restaurant will collapse ... Even if when family members died, I couldn't attend their funeral ... The recruiter even refuse to arrange an annual medical check-up for us.''}- (P12).

\end{quote}

\setlength{\parindent}{0cm} \paragraph*{Alleviating the Disadvantaged Situation of Ageing Job Seekers}

Compared with young people, the ageing people has two main concerns towards work: 1). Decreased physical fitness, which makes them very sensitive to workload (e.g., frequent breaks at work). 2). Family burden. Ageing females in China often need to take care of their families (e.g., young grandchildren), which prevents them from working long hours in the workspace. We find that companies that have made it clear that they are willing to accept ageing job seekers on video platforms tend to be more tolerant of ageing workers and understand their psycho-physical difficulties. Many of the jobs presented in the videos and live-streams are handcraft work (e.g., assembling toys in toy factories), which have particularly low requirements for physical fitness and skill levels. Generally, recruiters tend to present the daily tasks of jobs in videos, which are often a few simple fixed actions. These contents greatly motivate female ageing job seekers, who may have been away from work for a long time, to apply for the jobs. For example, some manufacturers offer female job applicants the opportunities to work from home while caring for their families. These part-time jobs allow ageing women to earn extra money in their spare time, e.g.,

\begin{quote}

\emph{``I found a pen assembling job from home through \emph{TikTok}. The work is paid by the amount of pens I assemble, and I get 0.1 yuan for each pen ... When you have assembled a sufficient number of pens and sent them back to the factory, you will be rewarded better.''} - P4.

\end{quote}

Through the access provided by short video entertainment platforms, disadvantaged ageing seekers may form groups and strive for more benefits from recruiters. P11 has two school-aged children that she must care for at home. She has indeed managed to build a handicraft team with some local women who also care for their families at home and is negotiating with the manufacturer for the restoration of their homemade fabric crafts. She is confident that if they take this opportunity and operate as a team, they will attract the attention of factories and gain more revenue. Cost-effectiveness is also a significant empowerment for ageing job seekers with unstable income. Compared with traditional offline job-seeking methods (e.g., offline employment agencies) they used in the past, seeking jobs through short video and live-streaming platforms is a highly cost-effective way to find a job: 
\emph{``Some labor agency want me to pay application fee in advance ... The application fee could be several hundred yuan ... The current job I found through TikTok don't need me pay any application fee.''} (P6) and \emph{``I need to travel very far to the labor agency ... I need to take the bus for forty minutes ... It also often takes me fifty minutes to wait for the bus.''} (P8). Furthermore,the expensive application fees charged by agencies are often non-refundable, even if the agency failed to find suitable positions for ageing job seekers. This highlights the urgent need for job-seeking methods that impose less financial pressure on ageing job seekers.

\section{DISCUSSION}

These results provide a nuanced understanding of the practices and experiences of ageing job seekers on
short video and live-streaming platforms. Our observations and findings complement prior research on job searching needs of disadvantaged groups, which highlighted the role of youth-oriented platforms
for ageing job seekers with limited digital literacy. We now reflect on the digital platform acceptance for 
ageing job seekers, the well-being of ageing workers in non-Western countries, and the collaboration with short-video and live-streaming platforms on job-seeking for ageing job seekers. We further offer several design implications to make short-video and live-streaming platforms more suitable for ageing job seekers to conduct job-seeking.     


\subsection{Rethinking Digital Platform Acceptance for the Ageing Job Seekers: From Learning Difficulty to Trust Building}


Our work reveals the overall job-seeking process of ageing people on short video and live-streaming platforms. Similarly, prior researchers in HCI have discussed the motivations and practices of the ageing to participate in online gig work through specialized digital platforms \cite{brewer2016would,skorupska2018older}. However, job-seeking through specialized job-seeking platforms is a more complex and serious task for our interviewees, which requires them to conduct a series of searching/filtering operations as well as identifying reliable recruiters before conducting on-site visiting. Despite help from family members, most of our interviewees still give up learning specialized job-seeking platforms. At the same time, some of our interviewees are suspicious of these platforms due to the lack of understanding of their operations. Tawanna emphasized the importance of trust for ageing individuals to search jobs on digital platforms \cite{dillahunt2018designing}. She also proposed that the perceived behavioral control could influence the intention of ageing individuals to seek employment, considering the challenges they face when using digital job-seeking platforms \cite{dillahunt2018designing}. Consistent with her findings, our interviewees tend to be highly resistant to specialized job-seeking platforms. Compared to learn a complex apps from scratch, introducing some useful and well designed mechanisms to entertainment platforms they have already easily use (e.g., insert fast application submission plug-in into live-streaming channels) is a better way to help them utilize digital platforms for job-seeking. Our interviewees could easily utilize these short video and live-streaming platforms to conduct job-seeking since most of them have spent a lot of time on these apps for entertainment.  Our findings also revealed that cost and time efficiency may serve as important motivations for older job seekers to incorporate digital platforms into their job-seeking process. For ageing job seekers without stable income, these familiar short video and live-streaming platforms may be proper choices for them to enjoy these benefits and minimize the learning difficulties they worry about. On the contrary, the complex digital platforms may impose a high learning burden on them while offering extremely minimal assistance. In summary, the limited digital literacy of ageing users presents a challenge in fully experiencing the benefits that many digital platforms were originally designed to provide for their users.   

\quad Prior work has shown that ageing job seekers who are not technologically literate (e.g., ageing return citizens) always avoid using social media or are highly resistant to finding positions through social media \cite{dillahunt2018designing}. This may be because of their lack of understanding and perceived usefulness for complicated systems \cite{boyd2015easisocial}. Interestingly, even without certain operational skills (e.g., searching and typing on mobile phones), our interviewees still tried and succeeded in finding suitable positions through short-video and live-streaming platforms. Similar to groups lacking understanding of social media \cite{page2022perceiving}, our interviewees can perceive unique affordance towards social media.Compared to regular text-based job-seeking apps, we also found that ageing people are more willing to trust and accept digital platforms providing visual presentations. Through video/live-streaming, they could be more clear of the details about the positions they are pursuing (e.g., the working environment). In previous work, researchers have proposed that distrust was frequently used by the ageing to justify their non-use of digital platforms for ageing people \cite{knowles2018wisdom,knowles2018older}. To construct the trust between digital platforms and the ageing, developers devote to improve the transparency and address misconceptions of the systems for ageing people \cite{frik2019privacy}. Our work is a great example of how videos-based system could be a potential solution to increase system transparency, which allows ageing users to directly understand the recruitment information and easily interact with system (e.g., question the recruiters in the live-streaming channels). We also found that the public regulations of the platforms may help the ageing build a stronger sense of trust towards the digital platforms. By forcing recruiters to disclose significant information (e.g., business certificate) and publicly promise monthly salary/welfare, ageing job seekers are more willing to submit applications. Previous studies have shown that younger job seekers utilizing online platforms tend to actively build their own online networks \cite{stone2019influence} and proactively search for job information on these platforms \cite{wadhawan2018factors}. However, our research indicates that ageing job seekers are more cautious about the risks associated with online job-seeking, often driven by their fear of potential scams and a general distrust of digital platforms. Further exploration is needed to understand the additional factors that contribute to the differences between ageing and younger online job seekers.

\subsection{The Well Being of Ageing Workers and Job Seekers in Non-Western Countries}

Researchers in HCI have discussed a lot concerning the well-being sense of workers, which encompasses their satisfaction with physical, financial, and psychological conditions \cite{lee2021participatory,mutari2015just}. Previous work has proposed that marginalized workers' well-being is more likely to be undermined. For example, You found that gig workers may incline towards long-hour work due to low hourly wages \cite{you2021go}. This result is strongly correlated with our findings: As their physical condition declines, ageing job seekers may become more vulnerable to severe inequalities, which can lead to difficulties in their employability and 
even survival. Previous HCI research has reported a similar issue occurring in the technology industry in Western countries; for instance, IT companies in the US tend to prefer younger candidates over more experienced ageing candidates \cite{hawthorn2000letter}. Age discrimination is not limited to the technology industry; we find that age discrimination spans across various industries, as reflected in the diverse industries where our interviewees were located. Previous HCI work has also discussed and observed the marginalized position of individuals in the labor market due to negative experiences, such as crime, vagrancy, and disability \cite{ogbonnaya2018returning,hendry2017homeless,mcewin2017working}. However, in our study, we found that even in the absence of these negative experiences, ageing workers are vulnerable to irreversible marginalization as they naturally age. Our interviewees were generally caught in the dilemma of facing skepticism from recruiters and being forced to compete with younger candidates on unequal terms. To support these ageing job seekers who are sidelined in the daily job market, we need to introduce special resources and tools (e.g., short videos and live-streaming) to help them find appropriate opportunities.

\quad Our results also highlight how short videos, as well as live-streaming channels, can serve as special recruitment tools to help ageing job seekers alleviate their disadvantaged situations and enjoy the basic welfare and rights they deserve. This aligns with previous work in HCI that explores the use of technology by disadvantaged groups to overcome job searching difficulties \cite{wheeler2018navigating}. Previous researchers have also suggested that digital platforms could positively impact workers' wellbeing \cite{nazareno2021impact}. However, ageing individuals have not been the primary focus of previous studies. This could be because the ageing unemployed population in need of jobs is more common in developing countries like China. Furthermore, we discovered that these platforms enable ageing women to re-enter the workforce. However, the majority of jobs available to ageing workers on these platforms are low-end blue-collar occupations. Although these positions involve simple and repetitive tasks, they do not offer clear career advancement opportunities. With the rapid advancement of technologies such as artificial intelligence and automated robotics, there is concern that these jobs may become obsolete in the near future.

\subsection{Design Implications}

In this section, we will report design implications to improve current short video and live-streaming platforms for ageing job seekers, which includes specialized job seeker mode, career support, and better distinction between recruitment content and normal content.

\subsubsection{Job Seeker Mode in Platforms}

Currently on both platforms, job seekers and general recreational users rely on the same recommendation algorithm and adhere to the same privacy policy. However, our interviewees generally expressed a desire for more job-related content and did not want their personal information, which they had made public to increase their chances of response, to be seen by anyone other than the recruiters. When designing short-video and live-streaming platforms, we need to take different needs of ageing job seekers and general entertainment users towards the platforms into account. For example, the platform could allow ageing job seekers to enter a dedicated job seeker mode and make the content uploaded to the user's home page in this mode only available to the platform's certified recruiters. In addition, this mode should allow the platform to actively push only the job related videos and channels to ageing job seekers, which improves the efficiency of their job searching process.

\subsubsection{Career Support For Ageing Job Seekers}

We found that the jobs available on the platforms were mainly blue-collar jobs. We would like to see the platform offer a wider variety of jobs, and provide ageing job seekers with content that will help them improve their professional knowledge and broaden their career paths according to their career goals, abilities and education levels.

\subsubsection{Distinguish Recruitment Content and Normal Content}

Although both platforms have some vetting of recruiters, ageing job seekers are sometimes unable to distinguish between regular content creators (e.g., normal streamers) and the recruiters who have been vetted by the platforms, which makes it possible for them to be scammed. We believe that there should be a clearer distinction between recruitment content and general content in platforms. For example, there can be special alerts when joining a recruiter's Live-streaming  channel, which allows ageing job seekers to distinguish whether the recruiter in the channel is reliable. Platforms should also conduct real-time censorship of regular live-streaming  or short video content, and automatically alert ageing job seekers when keywords such as ``recruitment'' appear, to prevent them from being deceived. However, it is worth noting that these platforms should maintain simple operations and familiar interfaces to avoid increasing the burden of relearning for ageing users. Our findings indicate that ageing users are highly sensitive to the learning curve associated with digital platforms. Platform designers should exercise caution when considering increasing the complexity of digital platform operations, as it may have an unexpected negative impact on the experience of ageing users. 

\subsection{Limitations and Reflections}

This work has several limitations. First, our study is based on a limited set of ageing interviewees, which may limit the generalizability of the proposed design implications to other contexts. However, we reached data saturation during our analysis, and our findings provide insights into the experiences of the interviewed ageing job seekers and observed content. We suggest that future research expand the interviewee pool and collect additional statistical information to enable a more comprehensive analysis.

\section{CONCLUSION}

\noindent This study on ageing people seeking jobs on short-video and live-streaming platforms reveals the injustice they suffered in their daily lives as well as how these platforms could help them alleviate this disadvantaged situation. Through a qualitative analysis of interviews and observational data, we pinpointed their motivations as a combination of dissatisfaction with offline labor market and specialized online job-seeking platforms. More notably, live-streams and short videos bring them more reliable job information and better protection of their basic welfare and rights. These merits, taken together, considerably alleviate the age discrimination suffered by ageing job seekers. In short, our depiction of ageing job seekers in short-video and live-streaming  platforms revealed that these platforms could potentially become a reliable and ageing-friendly access for ageing people to conduct job-seeking.

\bibliographystyle{ACM-Reference-Format}
\bibliography{sample-base}

\end{document}